\documentclass[11pt,twoside]{article}
\usepackage{asp2010}

\resetcounters

\begin{document}

\title{Imaging the Moon II: Webcam CCD Observations \& Analysis (a two week lab for non-majors)}
\author{Takashi~Sato
\affil{Department of Physics, Kwantlen Polytechnic University, 12666 72nd Ave, Surrey, BC, V3W 2M8, Canada}}

\begin{abstract}
Presented is a 
successful two week 
lab involving real sky observations of the Moon in which students make telescopic observations and analyze their own images.  Originally developed around the 35 mm film camera as a common household object adapted for astronomical work,  
the lab transitioned to use the webcam 
as film photography evolved into an obscure specialty technology and increasing numbers of students had little familiarity with it.  
The printed circuit board with the CCD 
is harvested
from a retail webcam 
and affixed to a tube to mount on
a telescope in place of an eyepiece.
Image frames are compiled to form a lunar mosaic
and crater sizes are measured.
Students also work through the logistical steps of telescope time assignment and scheduling, keeping to schedule and working with uncertainties of weather,  in ways paralleling research observations.  Because there is no need for a campus observatory, this lab can be replicated at a wide variety of institutions.
\end{abstract}

\section{Introduction}
Kwantlen Polytechnic University (KPU) offers an `Astro 101' 
survey course for non-majors, with weekly labs.
The activities presented here form a two part lab 
in which students make real sky observations in Week 1 and perform data analysis in Week 2.
Prior to this lab, students will have worked through a number of 
labs including basic optics and image formation with lenses and mirrors.

This lab aims to demonstrate 
that observational astronomy involves quantitative data acquisition with instrumentation 
and to 
demystify the imaging process by making observations using familiar, everyday equipment where possible, 
rather than 
specialized apparatus.
The lab also 
provides students opportunities 
to experience the joy and frustration of observing, and 
to make measurements from their own images.
The value of prior planning and good note-taking in scientific work is reinforced because of the requirement to work through complex tasks separated 
by time 
across two sessions.
The Moon has been selected as our real sky object for 
its large angular size and brightness.

An earlier version of this lab using 35 mm film was presented at 
\textit{Cosmos in the Classroom}
\citep{Sato_2004}  and was in use until 2009.
As photographic film became increasingly cumbersome to source, 
we recognized that film photography had moved out of the typical student's experience base.  
In the spirit of using familiar, everyday equipment, 
the webcam based imager my department experimented with in the 1990s (`KPU Mk I') was revisited.
At that time, the webcam was a new consumer technology and the user experience was not found to be sufficiently smooth, prompting us to defer 
classroom implementation.

\section{Equipment}
This
lab is designed 
for implementation without 
need for 
a campus observatory.
The main components 
are 
(1) a small telescope,
(2) a USB operated camera adapted from a webcam and
(3) a computer
(Figure \ref{camera}).
All are widely available consumer equipment.
At KPU, we use 8 inch telescopes which are larger than needed for lunar 
observations.

\articlefigurethree{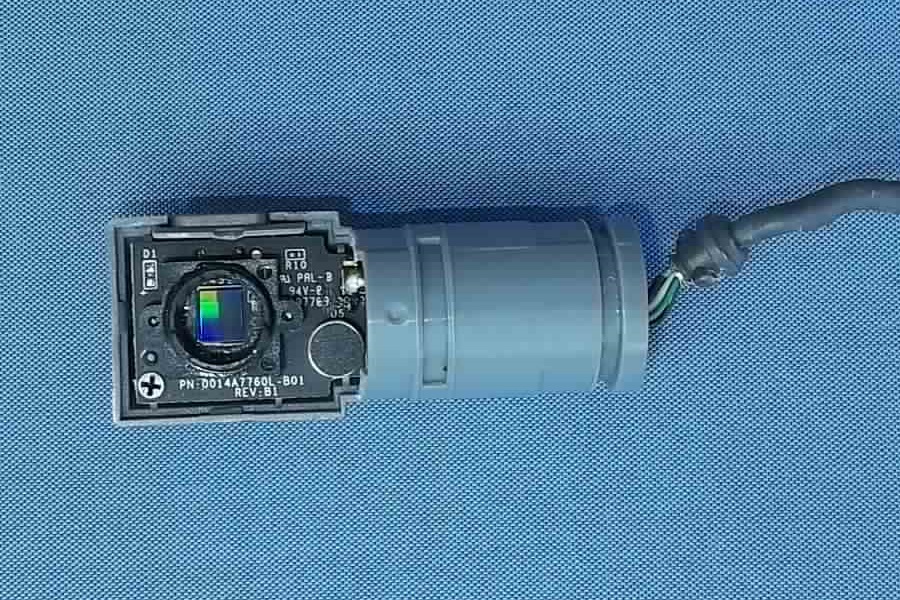}{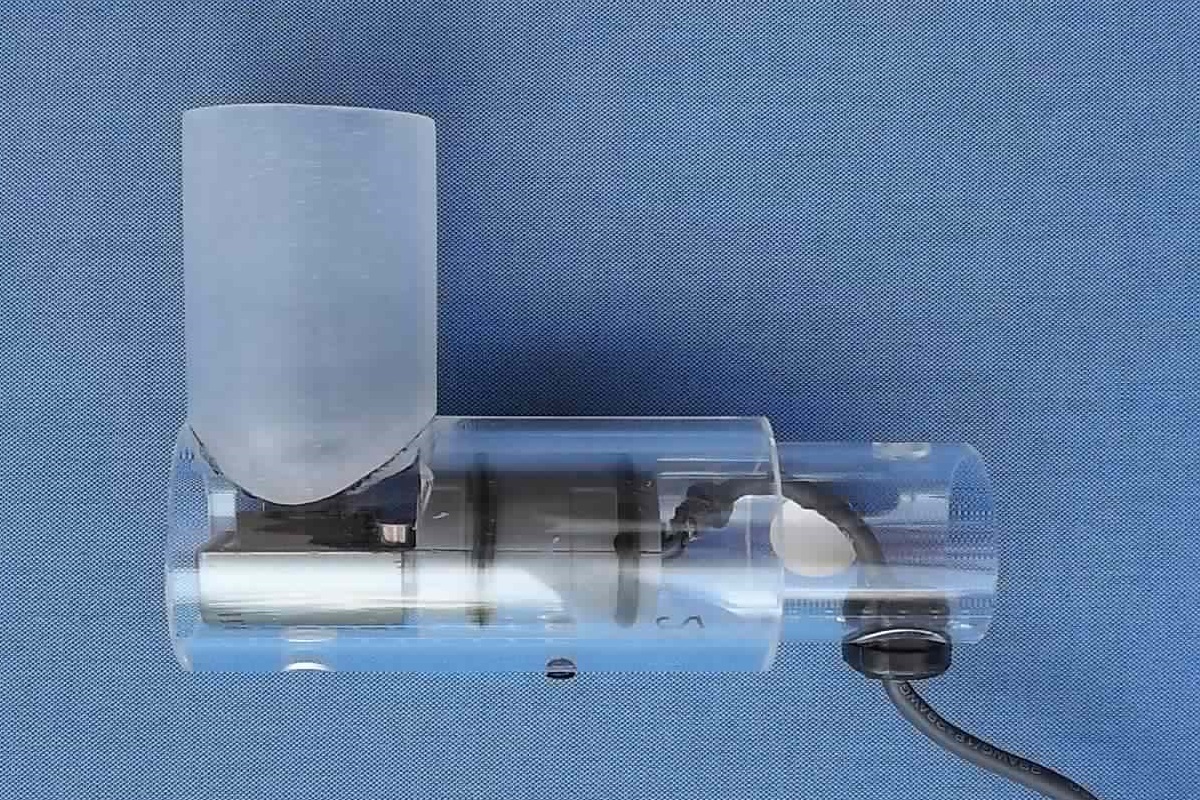}{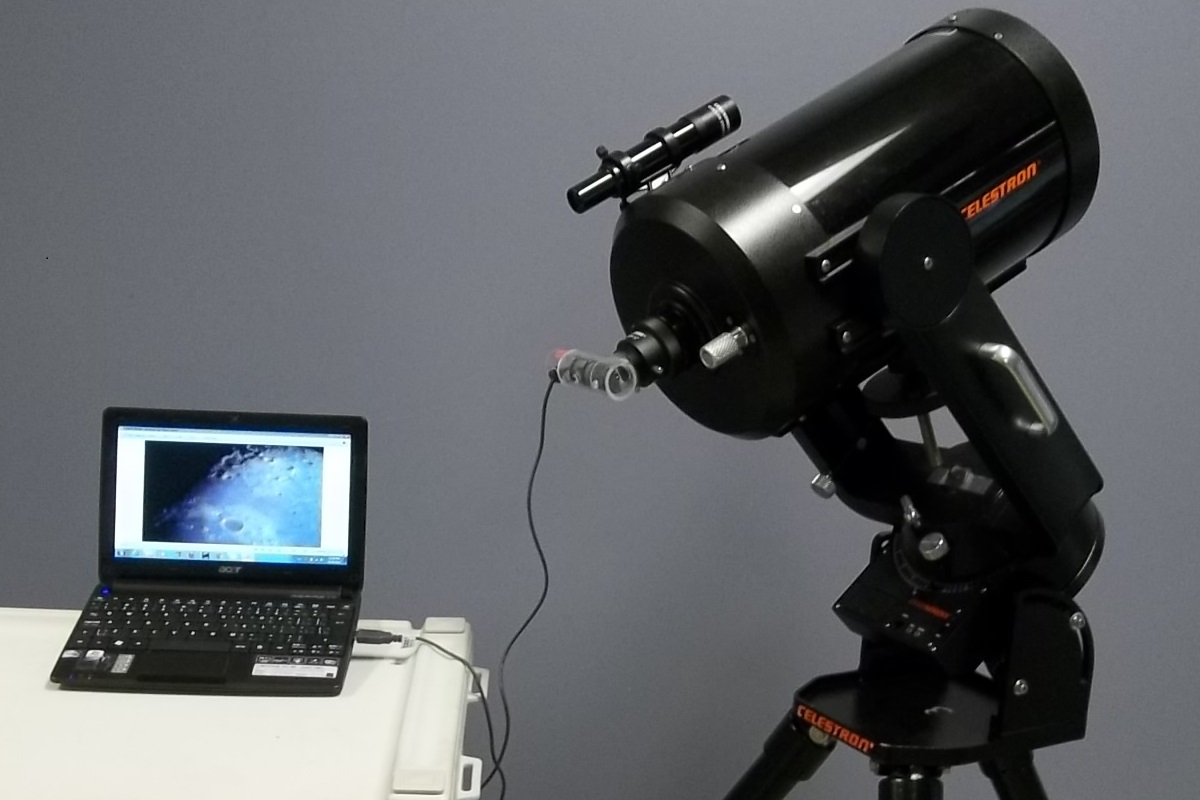}{camera}{Camera and telescope.
\textit{Left}: The lens and much of the casing is removed from an ordinary webcam.
\textit{Center}: KPU Mk II camera. The new packaging fits the eyepiece holder of telescopes.
\textit{Right}: An 8 inch telescope equipped with a KPU Mk II camera controlled by a netbook computer.
}

The adaptation and use of webcams is now well established in the amateur astronomy community 
\citep[e.g.][]{quickcam_buchanan,SchroderLuthen_2009}.
In this application, the lens  and much of the factory casing from a webcam 
is removed and discarded
(Figure \ref{camera} \textit{left}).
The remainder of the electronics is retained as-is.
A tube matching the diameter of an eyepiece is placed over the imaging chip such that the CCD can be placed directly at the focal plane of the telescope
(Figure \ref{camera} \textit{center}).
Although it is now possible to purchase an equivalent finished product from amateur astronomy retailers (e.g. Orion Starshoot USB Eyepiece Camera II) we undertook to produce our own cameras (KPU Mk II) to 
reinforce the idea that they are made from common webcams.

\section{Lab Activity}
At KPU, this lab runs over two sessions; one for observations and one for data analysis. 
After identifying ranges of dates with an evening Moon, 
groups of four students sign up for 50 minute observing shifts at our 8 inch telescopes.  
The class then stands by for the first clear night.
(Student lab manuals and related materials are available on-line.\footnote{See  http://www.kpu.ca/science/physics/faculty/takashi\_sato/COSMOS2013.html})

\subsection{Week 1 - Observations}
The first 30 minutes of the observing session is applied to orientation. 
Students 
make naked eye and eyepiece   observations to become familiar with the object to be imaged,
plan their grid patterns for mosaic construction,
and
prepare observing logs in their lab notebooks.
Each group is given a camera and a netbook computer.
They can also observe an earlier group taking data.

For the final 20 minutes, the group takes charge of 
a telescope and mounts its camera. 
A series of images are captured using stock webcam software.
As we do in research observations, students are prompted to ensure they obtain enough data for meaningful analysis, within the alloted telescope time.
Experience shows this is ample time for groups to complete their observations and for the class to stay on schedule. 
 
\subsection{Week 2 - Analysis}

Students produce a mosaic of the Moon's face
(Figure \ref{mosaic})
using digital tools. 
The readily available MS PowerPoint works 
well for this purpose and is familiar to students.
\articlefigure[width=7.9cm]{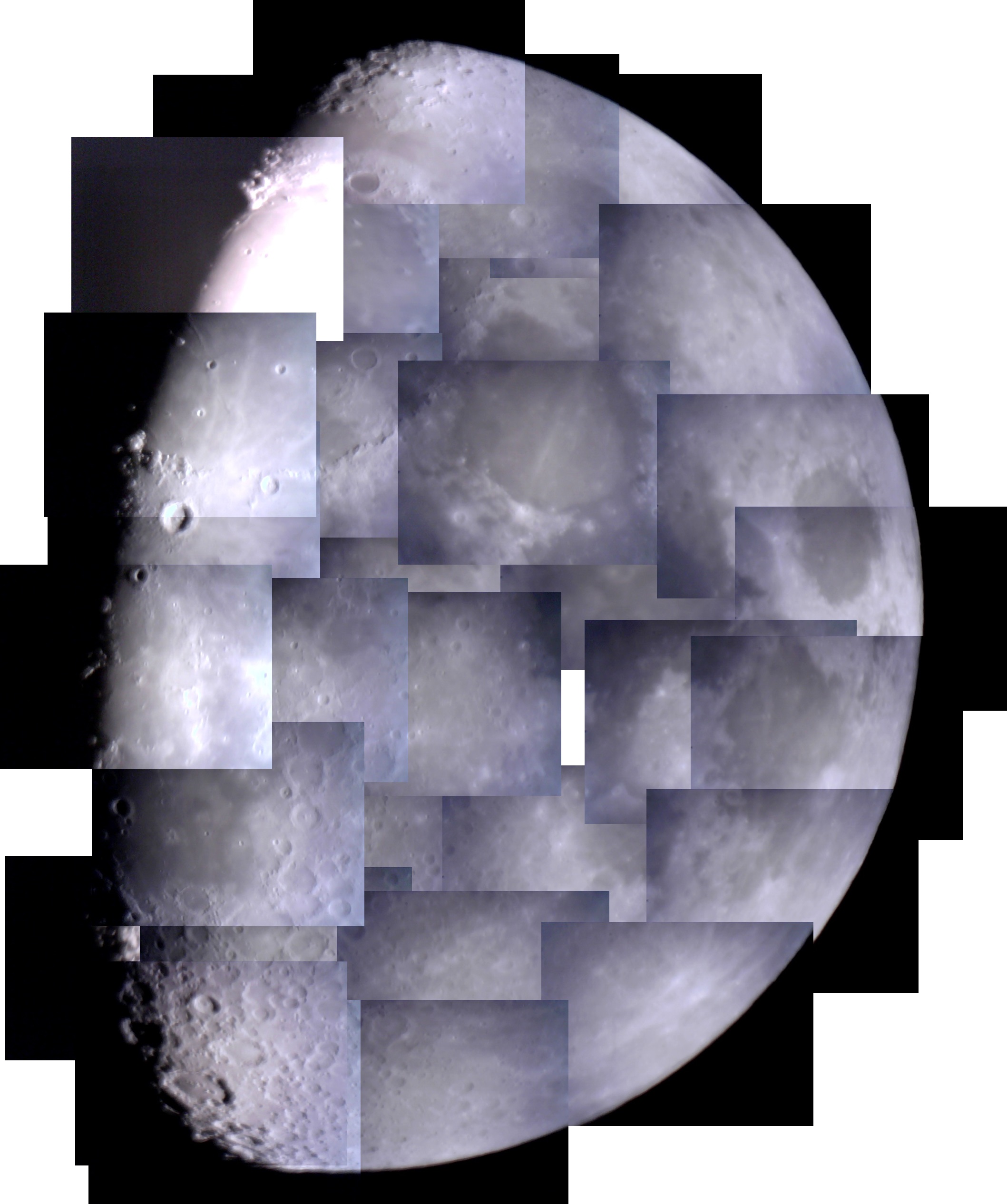}{mosaic}{Typical student result.
A mosaic is produced from a series of image frames.
The gap in coverage near the image center is a common and familiar feature.
The overexposed frame in the upper left is 
discussed in
Section \ref{technicalissues}.
}

Crater size measurement is a quantitative analysis
activity at a level 
appropriate for `Astro 101.'
Craters are identified 
on students' own images using lunar maps and globes.
Students then establish the image scale, measure crater sizes on the image and convert them to projected distances on the lunar surface. 
Students conclude by comparing each measured crater to a familiar geographical feature on Earth (e.g. ``crater Copernicus is about the size of Lake Erie").

\section{Discussion}

\subsection{Lab Development}
Simply replacing the original 35 mm camera with
an unmodified webcam results in an
image scale that makes it 
simple to acquire a very pleasing image of the Moon with only a single exposure,
making this lab trivial.
By removing the lens from the webcam, 
the new image scale places each frame across 
a small region of the Moon.
In order to build a mosaic, students must (1) plan their observations and
(2) take good notes (observing log). 
These steps align well with the learning objectives of the course.

\subsection{Technical Issues}
\label{technicalissues}
The webcam software used is furnished with the computer or the camera.  
By default, typical software appear to set exposure based on average pixel value across the frame.  This normally yields good results if the Moon fills a frame.  However, if the blank sky is a significant part of a frame, 
this algorithm is not optimal and 
the result is an overexposed lunar surface on gray background 
(see, for example, Figure\space\ref{mosaic} upper left). 
This was a known issue from the 1990's work with the KPU Mk I.
Although today's webcam software typically provides a manual override option, 
this refinement is considered beyond the scope of our non-majors
course and we simply retain the default settings.

An interesting and unexpected issue is that the CCD chips inside nominally identical cameras can vary in size, resulting in varying sky coverage.

\subsection{Student Work}
The resulting images 
(e.g. Figure \ref{mosaic}) display reasonably high quality, generally limited by seeing.  
Although software exists to assist in mosaic building and smoothing the frame edges, we consider this refinement also beyond the scope of this lab.

The mosaic building process has proven to be a good
illustration of the merits of good note taking in
 science.
While most students plan reasonable grid patterns and keep observing logs as instructed, many of
those who ignore these steps have found themselves unable to construct the mosaic.

There have also been examples of student excellence.  
Some students discover that by switching the telescope tracking off, 
they can naturally let the Moon drift across the CCD
while they capture images. 
Unlike with the 35 mm cameras, most students 
arrive confident on camera operation, 
so much so that we have not had to instruct them.

\section{Summary}
Imaging the Moon has been a successful lab implemented
with only modest equipment needs.
The observing sessions are fast-paced evenings, fulfilling for students and instructor alike.  
Students consistently report enjoying observing and learning that measurements can be made from their own images.

This lab has evolved to employ the 
common camera technology of the day 
(``no specialized apparatus'').  
In anticipation of a future without webcams, I 
have begun 
exploring the use of smartphones and look forward to making a report 
in the near future.

\acknowledgements I wish to thank
Mr. Bob Chin for the design and construction of the KPU Mk II cameras based on my prototype constructed using electrical tape.

\bibliography{sato}

\end{document}